\documentstyle[preprint,aps]{revtex}
\begin{document}
\draft
\title{Experimental Verification of the Gapless Point in the $S$=1 
Antiferromagnetic Bond Alternating Chain}
\author{M. Hagiwara$^{1,2}$, Y. Narumi$^{3,2}$, K. Kindo$^{2,3}$, 
M. Kohno$^{4}$, H. Nakano$^{4}$, R. Sato$^{2}$ \\ and M. Takahashi$^{4}$}
\address{$^1$The Institute of Physical and Chemical Research (RIKEN), Wako, 
Saitama 351-01, Japan \\ $^2$KYOKUGEN, Osaka University, Toyonaka 560, 
Japan \\ $^3$CREST, Japan Science and Technology Corporation (JST) \\
$^4$Institute for Solid State Physics, University of Tokyo, 
Roppongi, Tokyo 106, Japan}
\date{Received }
\maketitle
\begin{abstract}
Susceptibility and high field magnetization measurements have been 
performed on powder samples of an $S$=1 bond alternating chain compound 
[\{Ni(333-tet)($\mu$-N$_3$)\}$_n$](ClO$_4$)$_n$ (333-tet=tetraamine 
N,N'-bis(3-aminopropyl)-1,3-propanediamine).   As the temperature is 
decreased, the susceptibility exhibits a round maximum at around 120 K and 
decreases gradually down to 10 K, and then falls down rapidly with a 
logarithmic curvature which is behavior of the susceptibility of a 
gapless or a nearly gapless antiferromagnetic chain.   Magnetization up to
50 T at 1.4 
K shows no or a very small gap in this compound.  We have carried out 
numerical calculations for the 
$S$=1 antiferromagnetic bond alternating chain with various alternating 
ratios $\alpha$ and obtained a very good agreement 
between experiments and calculations for $\alpha$=0.6.   We verify
experimentally that the gapless 
point exists around $\alpha$=0.6.             
\end{abstract}
\pacs{75.10.Jm, 75.50.Ee, 75.50.-y}
Experimental and theoretical efforts on low dimensional magnetic systems 
have brought about a profound understanding of physics in many body 
quantum systems.  In particular, linear chain Heisenberg antiferromagnets 
(LCHA's) with the quantum spin number $S$=1 have been studied 
extensively \cite{renard,katsumata,hagiwara} in connection with 
Haldane's conjecture \cite{haldane} that there is an energy gap between 
the ground state and the first excited one for integer $S$, while not for 
half-odd integer $S$.    Recently, low dimensional 
oxide systems, such as inorganic spin-Peierls systems \cite{hase,isobe} and 
spin ladder systems \cite{azuma,dagotto}, have attracted a lot of interest.   

One of the recent interesting topics in $S$=1 LCHA's is bond 
alternation.    Using the Hamiltonian for the $S$=1 bond alternating  chain
 given by ${\cal H} = \sum_{i} \{1 - (-1)^i \delta \} 
\mbox{\bf S}_i \cdot \mbox{\bf S}_{i+1}$  where $\delta$ (0$\leq\delta\leq$1) 
is the parameter representing bond alternation and {\bf S}$_{i}$, {\bf
S}$_{i+1}$ $S$=1 spin operators, Affleck and 
Haldane~\cite{affleck} predicted 
that there is a massless (gapless) point 
at a certain alternating ratio $\delta_{c}$.  Subsequent numerical 
studies~\cite{singh,kato,yamamoto,totsuka,kitazawa,totsuka2} estimated the 
critical ratio $\delta_{c}$ as 0.25.    Moreover, it 
was shown numerically that the ground state is in the Haldane or the singlet 
dimer phase depending on $\delta$ (0.0$\leq\delta<$0.25 for the former and 
0.25$<\delta\leq$1.0 for the latter).    

In the recent years, some Ni bond alternating
compounds~\cite{vicente,coronado,escuer,escuer2} were 
synthesized and the magnetic properties were studied.    They were (a) 
$catena$-[\{Ni$_2$($\mu$
-N$_3$)$_3$(dpt)$_2$\}$_n$](ClO$_4$)$_n$ (dpt=bis(3-aminopropyl)amine), (b) 
\{Ni$_2$(EDTA)(H$_2$O)$_4$\}$_n$$\cdot$(2H$_2$O)$_n$ (EDTA=
ethylenediaminetetraacetic acid), (c) 
$trans$-[\{Ni(333-tet)($\mu$-N$_3$)\}$_n$](ClO$_4$)$_n$ (333-tet= 
N,N'-bis(3-aminopropyl)-1,3-propanediamine), (d)
[\{Ni$_2$(dpt)$_2$($\mu$-ox)($\mu$-N$_3$)\}$_n$](PF$_6$)$_n$
(ox=C$_2$O$_4$), and (e) 
[\{Ni$_2$(Medpt)$_2$($\mu$-ox)($\mu$-N$_3$)\}$_n$](ClO$_4$)$_n$ 
(Medpt=methyl-bis(3-aminopropyl)amine).   Alternating ratios $\alpha$ of the 
neighboring exchange constants were obtained from the fit of the 
susceptibility to a formula calculated by Borr\'{a}s-Almenar\cite{borras} using the Hamiltonian 
defined as,
\begin{equation}
   {\cal H} = J \sum_{i=1}^{N/2} ({\bf S}_{2i-1} \cdot {\bf S}_{2i}+\alpha {\bf 
   S}_{2i} \cdot {\bf S}_{2i+1}) . 
\end{equation}
However, no comparison between experiment and theory have been 
carried  out from the viewpoint of verification of the phase separation and the 
gapless point.   Note that the critical ratio $\alpha_{c}$ corresponding 
to $\delta_{c}$ is 0.6.    Magnetization curves at zero Kelvin 
of the $S$=1 LCHA with bond alternation were calculated with the 
Hamiltonian ${\cal H} = \sum_{i} [\{1 - (-1)^i \delta \} 
\mbox{\bf S}_i \cdot \mbox{\bf 
S}_{i+1}$+$g\mu_{\rm B}S_i^zH$+$D(S_i^z)^2$] where $g\mu_{\rm B}S_i^zH$ is the 
Zeeman term, $g$ the $g$ value of Ni$^{2+}$, $\mu_{\rm B}$ the Bohr 
magneton and $H$ the external magnetic field, and $D\sum_{i}(S_i^z)^2$ is
the single ion anisotropy term 
\cite{tonegawa}.   One of the important features in this study is 
that a 1/2-plateau appears in the magnetization curve at least when
0$<\delta\leq$1 and 
$D\geq$0.0.   Quite recently, the above finding was generalized and it 
was argued that the magnetization per site $m$ is topologically quantized 
as $n(S-m)=$integer at the plateaus, where $n$ is the period of the 
ground state\cite{oshikawa}.  Accordingly, in the 
$S$=1 antiferromagnetic bond alternating chain ($n=2$), $m$=1/2 can be 
quantized in addition to $m$=0 and 1.   Under such circumstances, some of
the authors
performed magnetization measurements on some of these compounds and 
observed 1/2-plateaus in magnetization curves of 
\{Ni$_2$(EDTA)(H$_2$O)$_4$\}$_n$$\cdot$(2H$_2$O)$_n$ 
and
[\{Ni$_2$(Medpt)$_2$($\mu$-ox)($\mu$-N$_3$)\}$_n$](ClO$_4$)$_n$~\cite{narumi
}.   

In the present paper, we report the results of susceptibility and high field 
magnetization measurements performed on powder samples of [\{Ni(333-tet)($\mu$
-N$_3$)\}$_n$](ClO$_4$)$_n$. We have found that the experimental results shows 
no or a very small gap in this compound and verified that 
$\alpha_{c}$ should be about 0.6 in comparison with some numerical 
results. 

First, we mention the synthesis and the crystal structure of
[\{Ni(333-tet)($\mu$
-N$_3$)\}$_n$](ClO$_4$)$_n$.   The compound [\{Ni(333-tet)($\mu$
-N$_3$)\}$_n$](ClO$_4$)$_n$ was synthesized with a slow evaporation 
method from an aqueous solution containing equimolars of 
Ni(ClO$_4$)$\cdot$6H$_2$O, 333-tet ligand and NaN$_3$ according to 
ref.\cite{escuer}.   This compound crystallizes in the triclinic 
system, space group $P\bar{1}$.   The lattice constants and angles at 
room temperature are $a$=8.765(1)\AA, $b$=8.976(1)\AA, $c$=11.995(2)\AA, 
$\alpha$=106.65(1)$^{\circ}$, $\beta$=110.06(1)$^{\circ}$ and
$\gamma$=91.11(1)$^{\circ}$.  Each Ni atom is placed in an octahedral 
environment and links to the neighboring Ni atoms via azido (N$_3$) groups to 
form the chain structure along the $c$ axis.  The chain structure of this 
compound is shown in Fig.~1.   Here, most of the ligand atoms and the 
counter anions ClO$_4^{-}$ are omitted for simplicity.   The neighboring 
chains are well separated from each other by ClO$_4^-$ anions.   Two 
kinds of alternating centrosymmetric azido bridges are present in the 
chain.   The Ni-N bond distances and Ni-N-N bond angles for these azido 
bridges are 2.077(3)\AA, 142.4(3)$^{\circ}$ and 2.204(3)\AA, 
123.6(2)$^{\circ}$, respectively, giving bond alternation in the 
chain.
Susceptibility ($M/H$) was measured at 100 Oe with SQUID magnetometer
(Quantum Design's 
MPMS2) in RIKEN and high field magnetization measurements were done with a
pulsed 
magnet up to 50 T at KYOKUGEN in Osaka University.     

Susceptibility of the powder sample of [\{Ni(333-tet)($\mu$
-N$_3$)\}$_n$](ClO$_4$)$_n$ is shown in Fig.~2.   A round maximum appears 
around 120 K and the susceptibility decreases gradually down to 10 K, and 
then it decreases steeply as the temperature is decreased further.   No 
long range order was observed down to 1.8 K.   Low 
temperature part (below about 30 K) of the susceptibility is quite similar
to that of the 
$S$=1/2 LCHA which was calculated recently by the Bethe ansatz 
and a field theory methods~\cite{eggert} and observed experimentally in an 
organic ion radical salt [3,3'-dimethyl-2,2'-thiazolinocyanine]-TCNQ 
\cite{takagi}.  From the similarity of these findings, we can say that this
system is 
gapless or nearly gapless because of a logarithmic fall down in the 
susceptibility at low temperatures   
\cite{moukouri}.   We compare the susceptibility with some numerical 
calculations (an exact diagonalization method and a quantum Monte Carlo
(QMC) method (loop algorithm)\cite{loop_alg}) for the 
Hamiltonian (1), including the single ion anisotropy term for the 
former case.   In the former calculation, the susceptibilities at and 
around the gapless points predicted in the $D$ versus $\alpha$ plane 
\cite{tonegawa} are calculated and compared with 
the experimental one.   Then, we conclude that the susceptibility can be 
fitted for a small $D/J$ ($<$0.1), $\alpha$ around 0.6 and $J/k_{\rm B}$
$\sim$ 100 
K. This small 
$D/J$ is reasonable, because Ni$^{2+}$ ion under an octahedral ligand field 
usually has the $D/k_{\rm B}$ value varying from 1 K to 10 K.    For 
$D/k_{\rm B}T \ll$ 1, 
the susceptibility for a powder sample can be compared with the 
calculated one for $D$=0.  In Fig. 2, open squares show the results of 
 the QMC method for 96 sites with $\alpha$=0.6.   The exchange 
constant $J/k_{\rm B}$ is estimated to be 110 K with the $g$ value of 2.46.   
The agreement between experiment and calculation is very good.   We show 
$\alpha$ dependence of the susceptibilities calculated by means of 
the QMC method (96 sites) in Fig.~3.   Here, we take the exchange constants
and the $g$ values 
to fit the experimental data at around 300 K.   The susceptibilities except
for $\alpha$=0.6 decrease 
exponentially at low temperatures.   It is easily mentioned from the 
calculated susceptibilities that the critical alternating ratio which 
corresponds to the gapless point exists around 0.6.   Details and further 
results of the Monte Carlo calculation will be presented in a longer
paper\cite{Kohno}.  

In order to confirm that this is a gapless or a nearly gapless system, we 
carried out high field magnetization measurements.   The result is shown 
in Fig.4 together with the result of a numerical calculation at zero Kelvin
made by the 
product-wave-function renormalization-group (PWFRG) method \cite{nishino}. 
The PWFRG method has been verified \cite{sato} to be more useful for
investigation 
of the magnetization process than the original density-matrix 
renormalization-group method devised by White \cite{white}.   The 
agreement between experiment and theory is also very good 
and the estimated exchange constant $J/k_{\rm B}$ is 115 K which is close 
to that obtained from the susceptibility fitting (110 K).    In the inset
of Fig.4, full magnetization 
curves up to the saturation field calculated by the PWFRG and a method 
introduced by Sakai and Takahashi (ST) \cite{sakai} are shown, together 
with experimental data.  According to the calculation, a 1/2 
plateau of magnetization curve, which is characteristic of the $S$=1 bond
alternating 
chains, should be expected around 140 T. 

In conclusion, we have performed susceptibility and magnetization 
measurements on an $S$=1 antiferromagnetic bond alternating chain compound
[\{Ni(333-tet)($\mu$
-N$_3$)\}$_n$](ClO$_4$)$_n$.   The susceptibility exhibits the behavior of 
a logarithmic fall down below 10 K, which is typical of the gapless 
antiferromagnetic chain.   By comparing experimental results with some 
numerical calculations for the $S$=1 bond alternating chain with 
$\alpha$=0.6, we have obtained a very good agreement and estimated the 
exchange constant $J/k_{\rm B}$=110 $\sim$ 115 K and the 
$g$ value of Ni$^{2+}$ ion $g$=2.46.   Consequently, we have verified 
experimentally that the gapless point exists around $\alpha$=0.6.

This work was carried out under the Visiting Researcher's program of
KYOKUGEN in 
Osaka University.   The authors thank the Supercomputer Center of 
Institute for Solid State Physics, University of Tokyo for the facilities 
and the use of the FACOM VPP500.   Part of the calculations were performed
on the Intel Japan PARAGON
at Institute for Solid State Physics, University of Tokyo.   Some of the
Computer Programs are 
based on ``KOBEPACK Ver.1.0'' coded by Professors T. Tonegawa, M. 
Kaburagi and Dr. T.Nishino.   This study was partially supported by a
Grand-in-Aid 
for Scientific Research from the Japanese Ministry of Education, Science, 
Sports and Culture.

 \begin{figure}
 \caption{Schematic view of the chain structure of [\{Ni(333-tet)($\mu$
-N$_3$)\}$_n$](ClO$_4$)$_n$. Most part of the 333-tet ligands and ClO$_4^-$ 
anions are omitted for simplicity.}
\label{fig1}
\end{figure}
 
 \begin{figure}
 \caption{Temperature dependence of the susceptibility ($M/H$) of a powder 
 sample of [\{Ni(333-tet)($\mu$-N$_3$)\}$_n$](ClO$_4$)$_n$ (filled
circles).  Open
squares show the calculated data by means of a quantum Monte Carlo (QMC) method 
for 96 sites for the $S$=1 bond alternating chain with the bond alternating 
ratio $\alpha$=0.6.}  
 \label{fig2}
 \end{figure}
 
 \begin{figure}
 \caption{Bond alternating ratio $\alpha$ dependence of the 
 susceptibility calculated with a QMC method (96 sites). }
 \label{fig3}
 \end{figure}
 
 \begin{figure}
 \caption{Magnetization of a powder sample of [\{Ni(333-tet)($\mu$
-N$_3$)\}$_n$](ClO$_4$)$_n$ at 1.4 K  versus magnetic field up to 50 T.  
Broken line shows a 
magnetization curve at 0 K calculated with the product-wave-function 
renormalization-group (PWFRG) method.   Inset: Calculated magnetization
curves up to the  
saturation field by the PWFRG method (broken line) and a method introduced
by Sakai 
and Takahashi (ST) (filled circles) , together with experimental data 
(solid line).}
 \label{fig4}
 \end{figure}
 
 \end{document}